\documentclass[aps,prl,twocolumn,superscriptaddress,showpacs]{revtex4-1}

\usepackage{amsmath}
\usepackage{graphicx}
\usepackage{calc}
\usepackage{bm}

\usepackage[T2A]{fontenc}
\usepackage[cp1251]{inputenc}
\usepackage[english]{babel}

\begin{document}

\title{ Thermoelectric response as a tool to observe electrocaloric effect in a thin conducting ferroelectric SnSe flake}

\author{N.N. Orlova}
\author{A.V. Timonina}
\author{N.N. Kolesnikov}
\author{E.V. Deviatov}

\affiliation{Institute of Solid State Physics of the Russian Academy of Sciences, Chernogolovka, Moscow District, 2 Academician Ossipyan str., 142432 Russia}

\date{\today}

\begin{abstract}
We experimentally investigate  thermoelectric response of a 100~nm thick SnSe single crystal flake under the current-induced dc electric field. Thermoelectric response appears as a second-harmonic transverse voltage $V_{xy}^{2\omega}$, which reflects temperature gradient across the sample due to the Joule heating by harmonic ac excitation current $I_{ac}$.  In addition to   strongly non-monotonous dependence  $V_{xy}^{2\omega}$, we observe that  dc field direction controls the sign of the temperature gradient in the SnSe flake.  We provide arguments, that electrocaloric effect is the mostly probable reason for  the results obtained. Thus, our experiment can be understood as  demonstration  of the possibility to induce electrocaloric effect by in-plane electric field in conducting ferroelectric crystals and to detect it by thermoelectric response. 
\end{abstract}


\maketitle

\section{Introduction}

Recent interest to conductors with broken inversion symmetry is connected not only with topological materials~\cite{armitage}, but also with conducting ferromagnetic and ferroelectric crystals. For ferromagnetic conductors with spin-orbit coupling, current-induced spin polarization leads to  spin-orbit torques~\cite{Jungwirth}, which opens a new field in spintronics. In ferroelectrics, current-induced electric field opens a way to control 
ferroelectric polarization in polar crystals. The latter effect can be expected for mono- or di- chalcogenides of transitional metals like WTe$_2$, SnS, SnSe, etc.~\cite{Lopez,Wte},  or for monolayer-based artificial structures~\cite{hBN_ferr,WS2MoS2}. 

One of the sophisticated phenomena in ferromagnetic or ferroelectric systems is the caloric effect, which is also important for applications. For example, it can be useful in development of new refrigeration technologies and materials~\cite{refrig,BTSn} and  cooling/heating environmentally friendly devices~\cite{cooling} or renewable energy sources~\cite{review,Ponomareva1}. Magnetocaloric and electrocaloric effects   occur due to the entropy difference between (ferromagnetically or ferroelectrically) ordered or disordered states, which can be controlled by applying or removing of the external magnetic or electric fields, respectively~\cite{ECE,BTSn}. For the electrocaloric effect, this difference leads to the  temperature variation if the ferroelectric polarization changes at stable entropy experimental environments~\cite{Ponomareva2,ferrECE}. Recent experimental investigations are performed for  insulating ferroelectric crystals~\cite{BTSn} or thin films~\cite{ECE,ferrECE,PZT}, which are placed between two metallic capacitor plates. Electrocaloric effect is controlled by external out-of-plane electric field in this case. In addition to general fundamental problems~\cite{Ponomareva1,Ponomareva2}, applied research is   mostly intended to improve the caloric effect in lead-free insulating materials~\cite{Pb-free1,BTSn}.

On the other hand, electrocaloric effect should principally be observable in ferroelectric conductors. Ferroelectric polarization is also  sensitive to current-induced electric field in conducting structures, which is impossible for  ferroelectric insulators. Even for small absolute values of the effect, corresponding temperature gradients should be detectable by in-situ thermoelectric response measurements~\cite{Esin,Esin1,Shash}. Among different materials, layered SnSe single crystals can be convenient for these investigations: thin SnSe flakes (300~nm and below)  are characterized by the in-plane ferroelectric polarization~\cite{FESnSe,SnSe} in addition to significant  thermoelectric properties~\cite{TE}.

Here, we experimentally investigate  thermoelectric response of a 100~nm thick SnSe single crystal flake under the current-induced dc electric field. Thermoelectric response appears as a second-harmonic transverse voltage $V_{xy}^{2\omega}$, which reflects temperature gradient across the sample due to the Joule heating by harmonic ac excitation current $I_{ac}$.  In addition to   strongly non-monotonous dependence  $V_{xy}^{2\omega}$, we observe that  dc field direction controls the sign of the temperature gradient in the SnSe flake.  We provide arguments, that electrocaloric effect is the mostly probable reason for  the results obtained. Thus, our experiment can be understood as  demonstration  of the possibility to induce electrocaloric effect by in-plane electric field in conducting ferroelectric crystals and to detect it by thermoelectric response.

\section{Samples and techniques}

\begin{figure}[t]
\center{\includegraphics[width=\columnwidth]{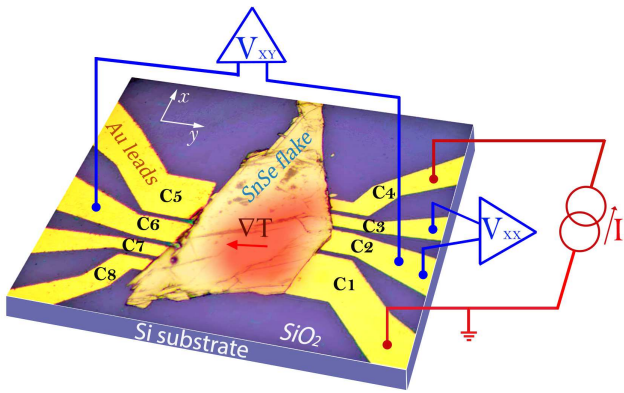}}
\caption{(Color online) Optical image of a typical sample with a sketch of electrical connections.   200 nm thick SnSe flake  is placed on the top of the pre-defined Au leads pattern, the 5~$\mu$m separated Au leads specifies experimental geometry for thermoelectric and charge  transport investigations. For thermolelectric measurements, thermal gradient $\nabla T $ is created by  ac current $\sim cos(\omega t)$ between two Au-SnSe contacts C1 and C4 due to the Joule heating of the sample, $\nabla T \sim cos(2\omega t)$ is perpendicular to the current line. Thus, we detect thermoelectric response as the second-harmonic transverse ac voltage $V_{xy}^{2\omega} \sim cos(2\omega t)$ between contacts C2 and C6.}
\label{sample}
\end{figure}

SnSe compound was synthesized by reaction of selenium vapors with the melt of high-purity tin in evacuated silica ampoules.  The SnSe layered single crystal was grown by vertical zone melting in silica crucibles under argon pressure. The structure of single crustal is verified by X-ray diffraction methods. The initial SnSe is characterized the layered structure with orthorhombic crystal system~\cite{SnSe}.

In layered monochalcogenides, ferroelectricity appears below some critical thickness~\cite{Lopez}, which can be estimated~\cite{SnSeprop} as 300~nm for SnSe.
 Ferroelectric polarization is due to the distortion of centrosymmetric $Pnma$ orthorhombic structure at low thicknesses,  which appears as polar orthorhombic $P2_1mn$ space group~\cite{FESnSe}.  In particular, anisotropy of the  Raman intensity is investigated vs. the SnSe flake thickness in  Ref.~\cite{SnSeprop}.
  The observed anisotropy has been connected with crystal symmetry group. It has been shown, that  SnSe symmetry is highest for the bulk material, while  unique in-plane anisotropic phonon behavior is observed for the  SnSe flake thickness from 300 nm  and down to the few-layer   samples~\cite{SnSeprop}.

Ultra-thin~\cite{SnSe} SnSe flakes (about 100-200~nm) are obtained by regular mechanical exfoliation from the initial layered ingot.  For electrical measurements the exfoliated flake is placed on the top of the pre-defined Au leads pattern, the pattern specifies experimental geometry for charge transport as it is depicted in Fig.~\ref{sample}.   Standard photolithography and lift-off technique are used to define  100~nm thick Au leads on the insulating SiO$_2$ substrate.  This procedure  has been verified  to provide electrically stable contacts with highly transparent metal-semiconductor interfaces, see Refs.~\cite{cdas,timnal}, and, simultaneously, it  minimizes chemical or thermal treatment of the initial flake~\cite{black}. It is also important, that the relevant (bottom) SnSe flake surface is protected from any oxidation or contamination by SiO$_2$ substrate.  

Resistance of the investigated samples varies  from 1~kOhm to 20~kOhm, an actual value depends mostly on the overlap area between SnSe flake and Au leads in Fig.~\ref{sample}.  

For thermolelectric measurements we use four-point lock-in technique with second harmonic detection~\cite{Esin,Esin1,Shash}. Thermal gradient is created by ac current $I_{ac} cos(\omega t)$ applied  between two Au-SnSe heating contacts C1 and C4, see Fig.~\ref{sample}, which we refer as $x$ axis. Thermal gradient appears due to the Joule heating of the sample $\nabla T \sim (I_{ac})^2 cos(2\omega t)$, it is perpendicular to the current line in Fig.~\ref{sample}. For this reason, we detect thermoelectric response as the second-harmonic transverse (i.e. along $y$ axis) voltage $V_{xy}^{2\omega}$ between contacts C2 and C6. To  obtain $V_{xy}^{2\omega}$ dependence on the in-plane electric field, we additionally apply high dc current $I_{dc}$ between the same heating contacts in Fig.~\ref{sample}. We wish to note, that longitudinal (along $x$)  dc current can not directly contribute to  the transverse (along $y$) $V_{xy}^{2\omega}$ second-harmonic ac response.

Amplitude and frequency of $I_{ac}$ is verified to have the correct Ohmic behavior of the longitudinal first harmonic $V_{xx}^{1\omega}$ component. In particular, $I_{ac}$  is below 10~$\mu$A  at  the frequency of 1.7~kHz. $I_{dc}$ is swept within $\pm$1~mA range, which corresponds to $10^9$~A/m$^2$ current density for our dimensions, and to in-plane electric field $10^5$~V/m for 1~kOhm sample resistance.  All measurements are performed at room temperature.

\section{Experimental results}
 
\begin{figure}[t]
\center{\includegraphics[width=\columnwidth]{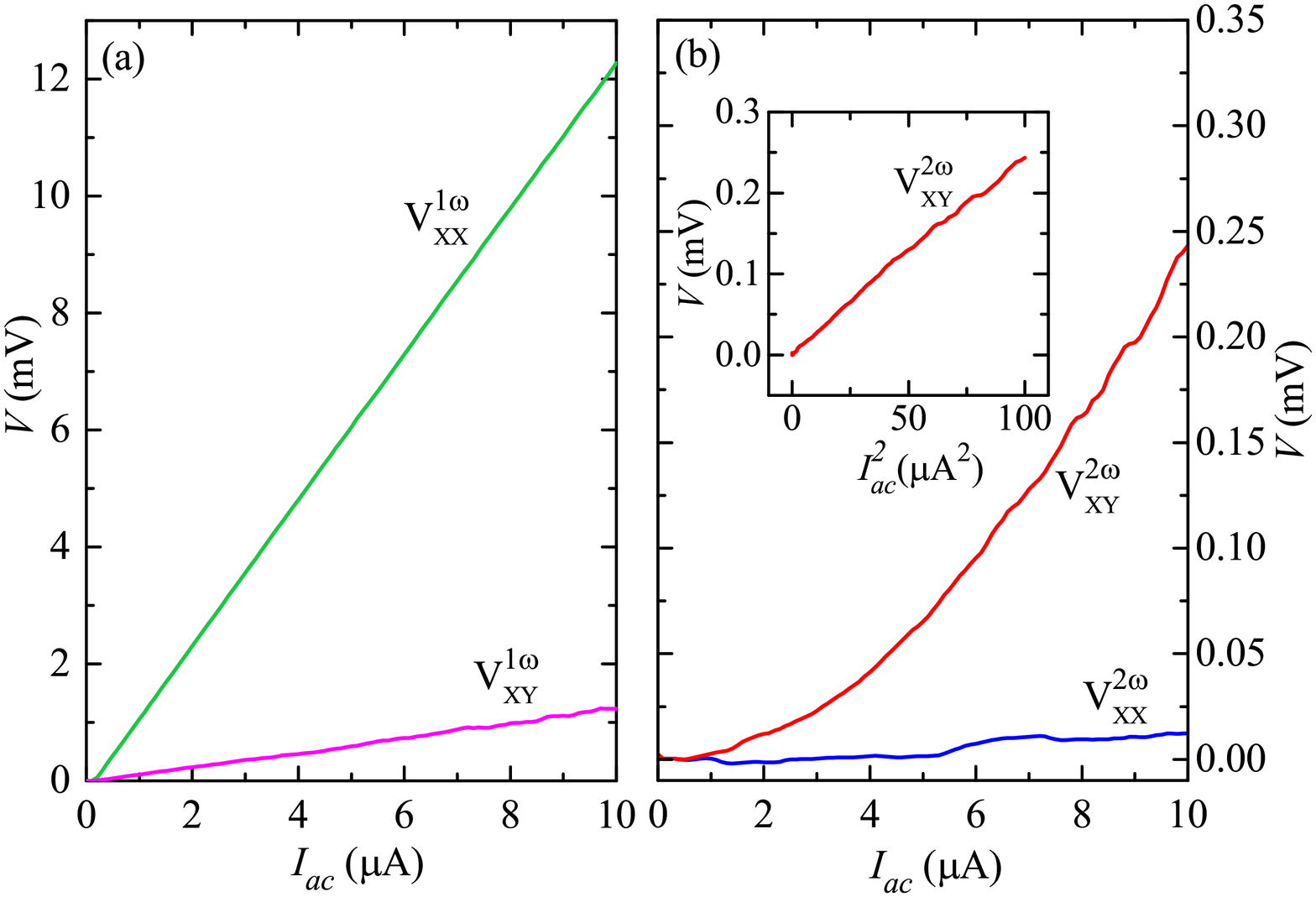}}
\caption{ (Color online)
(a) The first harmonic longitudinal voltage $V_{xx}^{1\omega}$ component in dependence on the ac current $I_{ac}$ amplitude. It demonstrates standard Ohmic behavior $V_{xx}^{1\omega} \sim I_{ac}$ with corresponding four-point sample resistance value  $R=$1.2~kOhm.  The first harmonic transverse voltage $V_{xy}^{1\omega}$ is much smaller, it seems to appear due to the contacts mismatch.    (b) Longitudinal  $V_{xx}^{2\omega}$ and transverse $V_{xy}^{2\omega}$ voltage components in dependence on the ac current $I_{ac}$ amplitude. $V_{xx}^{2\omega}$ is about zero, as it should be expected for the linear Ohmic $V_{xx}^{1\omega}(I_{x}^{ac})$ curve. In contrast,  there is significant  transverse second-harmonic voltage $V_{xy}^{2\omega}$, which is  non-linear and follows to $(I_{ac})^2$ law, as it is shown in the inset. This behavior well corresponds to thermoelectric origin of $V_{xy}^{2\omega} \sim \nabla T$. }
\label{VxxandVxy}
\end{figure}

Fig.~\ref{VxxandVxy} shows typical examples of four-point  longitudinal (along $x$) $V_{xx}^{1\omega}$, $V_{xx}^{2\omega}$ and transverse $V_{xy}^{1\omega}$, $V_{xy}^{2\omega}$ voltage components in dependence on the ac current  amplitude $I_{ac}$. The first harmonic longitudinal voltage $V_{xx}^{1\omega}$ component demonstrates standard Ohmic behavior $V_{xx}^{1\omega} = R I_{ac}$ with four-point sample resistance  $R=$1.2~kOhm, see Fig.~\ref{VxxandVxy} (a).  The first harmonic transverse voltage $V_{xy}^{1\omega}$ is much smaller, it seems to appear due to the contacts mismatch.   Ohmic behavior is also confirmed by nearly zero  second-harmonic longitudinal  $V_{xx}^{2\omega}$ component in  Fig.~\ref{VxxandVxy} (b). In contrast, there is significant  transverse second-harmonic voltage $V_{xy}^{2\omega}$ in Fig.~\ref{VxxandVxy} (b). $V_{xy}^{2\omega}$ it is clearly  non-linear and follows to $(I_{ac})^2$ law, as depicted in the inset to Fig.~\ref{VxxandVxy} (b). This behavior well corresponds to thermolelectric origin of $V_{xx}^{2\omega} \sim \nabla T$, where the temperature gradient $\nabla T$ is defined by Joule heating of the sample $\nabla T \sim (I_{ac})^2 cos(2\omega t)$. One can estimate maximum temperature difference $\Delta T$ as $\approx 0.5$~K between contacts C6 and C2 for the known~\cite{TE} SnSe thermoelectric coefficient 520$\mu$V/K.

\begin{figure}[t]
\center{\includegraphics[width=\columnwidth]{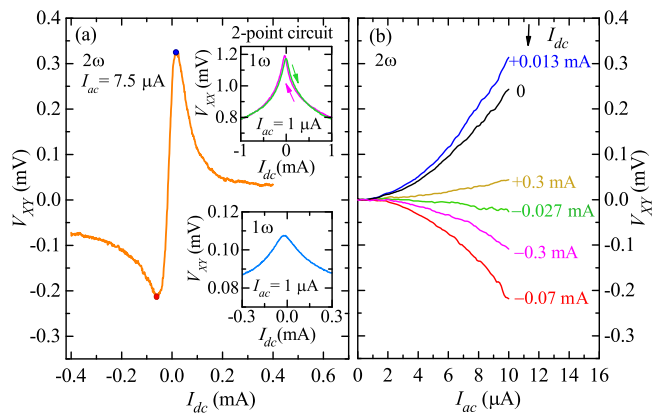}}
\caption{ (Color online) (a) Dependence of the thermoelectric response $V_{xy}^{2\omega}$ on the dc bias current $I_{dc}$, which is applied between the same contacts C1 and C4 as the ac current component. The  curve  is strongly non-monotonous, it consists of two   $\sim 1/I_{dc}$ branches with sharp switching between them around zero dc bias: $V_{xy}^{2\omega}$ is negative for $I_{dc}<0$, while it is positive for positive current values.  Upper inset demonstrates  first-harmonic longitudinal $V_{xx}^{1\omega}$ component, which reflects sample differential resistance.  Two curves are shown for two opposite sweep directions, which reflects ferroelectric hysteresis, see Ref.~\cite{wte2mem} for details.   Lower inset shows first-harmonic transverse voltage $V_{xy}^{1\omega}$. It differs significantly from the second-harmonic one in the main plot, and strongly resembles the xx component due to the contacts mismatch. 
(b) Square-type thermoelectric dependence $V_{xy}^{2\omega} \sim (I_{ac})^2$ for several fixed $I_{dc}$ values, $I_{dc} = +0.3; +0.013; 0; -0.027; -0.07; -0.3 ~mA$. The curves differ only by proportionality coefficient, which depends on the value and sign of $I_{dc}$ as $\sim 1/I_{dc}$. This behavior well confirms thermoelectric origin of $V_{xy}^{2\omega} \sim \nabla T$ at any $I_{dc}$ value, so it is the temperature gradient $\nabla T$ which depends as  $\sim 1/I_{dc}$ law.}
\label{Vxy}
\end{figure}

Our main result is the dependence of the thermoelectric response $V_{xy}^{2\omega}$ on the dc bias current $I_{dc}$, which is applied between the same contacts C1 and C4 as the ac current component. The experimental  $V_{xy}^{2\omega}(I_{dc})$ curve    consists of two   $\sim 1/I_{dc}$ branches with sharp switching between them around zero bias, see Fig.~\ref{Vxy} (a). Surprisingly, there is inversion of the thermoelectric response
sign with the direction of $I_{dc}$: $V_{xy}^{2\omega}$ is negative for $I_{dc}<0$, while it is positive for positive current values. 

We wish to note, that  the dc current contribution to Joule heating $\sim (I_{dc})^2$ is not sensitive to the current direction. Also, longitudinal dc  bias $I_{dc}$ can not be electrically detected  in the transverse ac second-harmonic responce  $V_{xy}^{2\omega}$. On the other hand, the odd antisymmetric second-harmonic $V_{xy}^{2\omega}(I_{dc})$ curve is in sharp contrast to  usual symmetric resistance behavior, which is shown as first-harmonic longitudinal $V_{xx}^{1\omega}$ in the upper inset to Fig.~\ref{Vxy} (a).  The first-harmonic transverse voltage $V_{xy}^{1\omega}$ is small, it qualitatively reproduces the $V_{xx}^{1\omega}$ behavior, as one could expect for small contact mismatch, see the lower inset to Fig.~\ref{Vxy} (a).

We check, that $V_{xy}^{2\omega}$ still reflects the sample thermoelectric response at any finite $I_{dc}$. Fig.~\ref{Vxy} (b) shows $V_{xy}^{2\omega} \sim (I_{ac})^2$ dependence for several fixed $I_{dc}$ values, the curves differ only by proportionality coefficient, which depends on the sign and value  of $I_{dc}$ as $\sim 1/I_{dc}$. Thus, the direction of $I_{dc}$ indeed affects the direction of the temperature gradient $\nabla T \sim V_{xy}^{2\omega}$, which can not be due to the dc current contribution to Joule heating $\sim (I_{dc})^2$.

\begin{figure}[t]
\center{\includegraphics[width=\columnwidth]{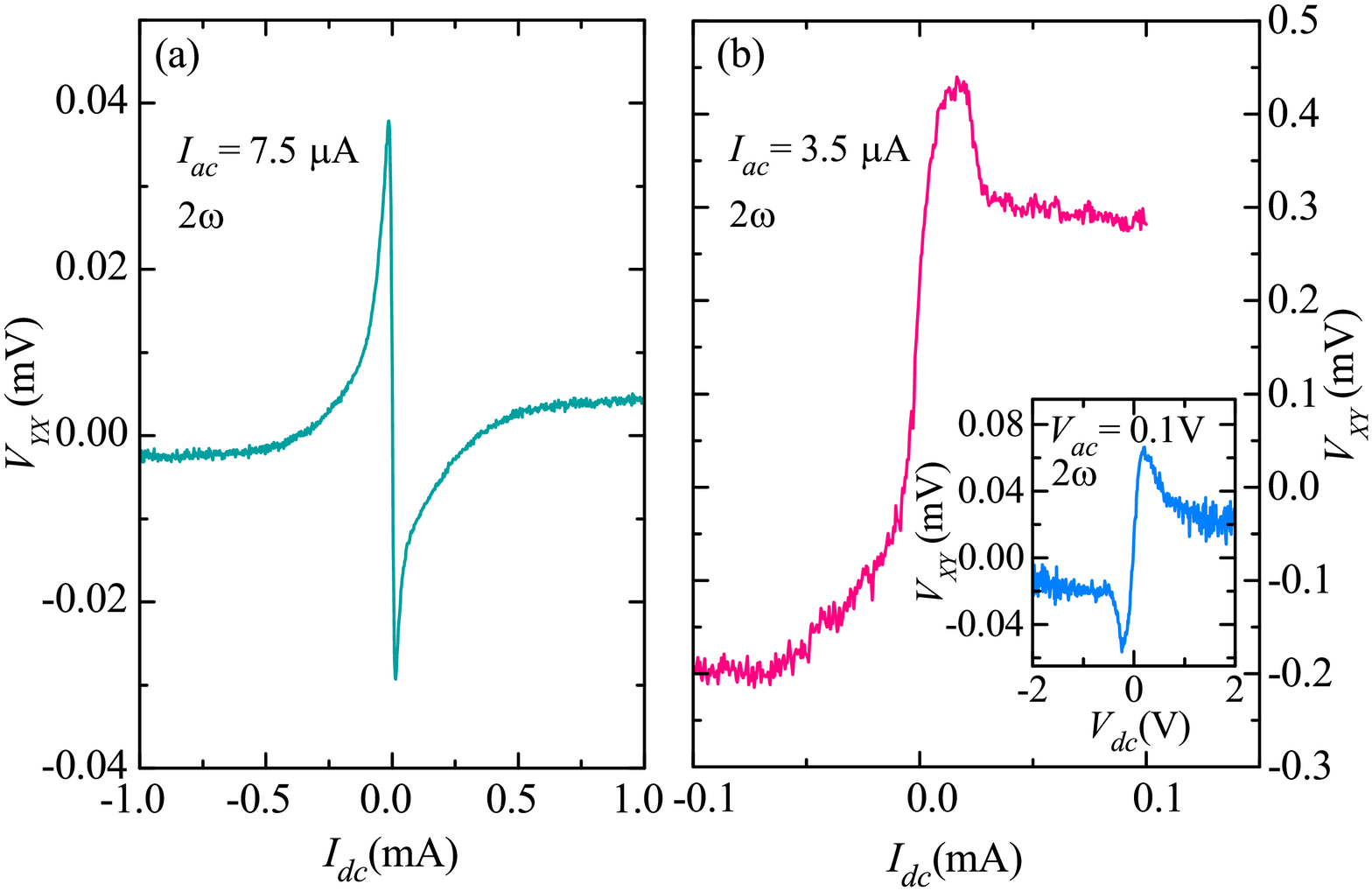}}
\caption{(a) Qualitatively similar  $V_{yx}^{2\omega} \sim 1/I_{dc}$ dependence for the  exchanged current and voltage probes in Fig.~\ref{sample}. Both current components $I_{dc}$ and $I_{ac}$ are applied along $y$ between contacts C2 and C6, while $V_{yx}^{2\omega}$ is measured between contacts C1 and C4 ($x$ axis). There are additional high-bias crossing points, however, $\nabla T$ sign inversion  around zero bias is similar to the the previous configuration. Thus,  antisymmetric behavior of $V_{xy}^{2\omega}(I_{dc})$ is not connected with any specific direction in the sample.
(b) Similar results for the sample with much higher resistance (about 20~kOhm), while the $I_{dc}$ range is narrowed in this case. Inset shows $V_{xy}^{2\omega}(V_{dc})$ curve with two $\sim 1/V_{dc}$ branches if the bias voltage is directly applied to the heating contacts in Fig.~\ref{sample}. Thus,  the sign of the theroelectric response is indeed determined by the direction of the in-plane dc electric field. 
}
\label{Vxy_II sample}
\end{figure}

The antisymmetric behavior of $V_{xy}^{2\omega}(I_{dc})$ is independent of the particular choise of contacts or any specific direction within SnSe flake. 
 Fig.~\ref{Vxy_II sample} (a) shows qualitatively similar  $V_{yx}^{2\omega} \sim 1/I_{dc}$ dependence for the  exchanged current and voltage probes in comparison to Fig.~\ref{sample}. In this case both current components $I_{dc}$ and $I_{ac}$ are applied in $y$ direction between contacts C2 and C6, while transverse $V_{yx}^{2\omega}$ is measured along $x$ between contacts C1 and C4, just opposite to the configuration in Fig.~\ref{sample}.  Thus, the antisymmetric behavior of $V_{xy}^{2\omega}(I_{dc})$ is  not connected with any specific sample inhomogeneity. 

These results can be also qualitatively reproduced for the sample with much higher resistance (about 20~kOhm). Fig.~\ref{Vxy_II sample} (b) shows antisymmetric odd  $V_{xy}^{2\omega}(I_{dc})$ dependence, while the $I_{dc}$ range is narrowed in this case. Due to the resistive sample, it is possible to directly apply bias voltage to the heating contacts in Fig.~\ref{sample}. The result is shown in the inset to Fig.~\ref{Vxy_II sample} (b) as antysymmetric $V_{xy}^{2\omega}(V_{dc})$ curve with two $\sim 1/V_{dc}$ branches, so the sign of the theroelectric response is determined by the direction of the in-plane dc electric field.

\section{Discussion}

As a result, we demonstrate that the transverse  second-harmonic ac voltage response $V_{xy}^{2\omega}$ indeed reflects temperature gradient $\nabla T$ at any $I_{dc}$ value. To our surprise, $\nabla T$ obeys $\sim 1/I_{dc}$ dependence with inversion of the $\nabla T$ sign with the direction of $I_{dc}$.

First of all, $\nabla T$ sign inversion can not be explained by simple geometrical factor.  In Fig.~\ref{sample} temperature gradient $\nabla T$ is measured between 80~$\mu$m spaced contacts C2 and C6, while the distance between heating contacts C1 and C4 is about 40~$\mu$m. Since the contact resistance exceeds the bulk SnSe value~\cite{SnSe}, Joule heating is mostly concentrated in the contact areas. Since two-pont resistance strongly depends on the dc bias in the upper inset to Fig.~\ref{Vxy} (a), the bias  changes relative contribution of the particular contact to  the Joule heating.  It leads to some variation of $\nabla T$ direction between C6-C1 and C6-C4 lines in Fig.~\ref{sample}, i.e. within $\approx \pm 15$ degrees, so it can not change $\nabla T$ sign in Fig.~\ref{Vxy} and in Fig.~\ref{Vxy_II sample} (b). In contrast, the curve in Fig.~\ref{Vxy_II sample} (a) is obtained in the alternative geometry, where voltage probes are situated to both sides from the current line. In this case, $\nabla T$ sign inversion is possible at  high dc biases, which seems to be responsible for the additional high-bias crossing points in Fig.~\ref{Vxy_II sample} (a).  However, $\nabla T \sim 1/I_{dc}$ dependence  around zero bias does not allow geometrical explanation also in this case. 

The first power of $I_{dc}$  indicates that $\nabla T$ is sensitive to the sign and value of the dc electric field. However, it can not be connected with the Peltier effect, since $\nabla T$ is proportional to  $\sim 1/I_{dc}$ rather than the expected  $\sim I_{dc}$ dependence for the Peltier effect. 

 On the other hand, electgrocaloric effect should principally be observable in ferroelectric conductors.  In contrast to standard ferroelectric insulator films~\cite{ECE,ferrECE,PZT}, it can be produced by in-plane current-induced electric field in conducting ferroelectric systems.

Recently, three-dimensional WTe$_2$ single crystals were found to demonstrate coexistence of metallic conductivity and ferroelectricity at room temperature~\cite{Wte} due to the strong anisotropy of the non-centrosymmetric crystal structure. The spontaneous polarization of ferroelectric domains was found to be bistable, it can be affected by high external electric field. The possibility to induce polarization current by source-drain field variation has been shown for WTe2  as  a direct consequence of ferroelectricity and metallic conductivity coexistence~\cite{wte2mem}. We have demonstrated qualitatively similar ferroelectric behavior of dV/dI(I) curves for thin  SnSe flakes~\cite{SnSe},   as it is shortly shown in the present text as small hysteresis  for  $V_{xx}^{1\omega}(I)$ in the upper inset to Fig.~\ref{Vxy}.

 Thin SnSe layers (300~nm and below) are  characterized~\cite{FESnSe} by in-plane spontaneous ferroelectric polarization at room temperature.  Ferroelectric domains are much smaller than the contact size in  our samples~\cite{FESnSe}:  the domains are  about 100 nm, the domain wall region is about 20-50 nm.   In this case, any variation of  the source-drain bias $I_{dc}$ affects ferroelectric polarization due to the domain wall shift for varying $E \sim I_{dc}$ in-plane electric field~\cite{SnSe}. It leads to  the  temperature variation because of the electrocaloric effect~\cite{Ponomareva2,ferrECE,defECE}, so the sign of the temperature $\delta T$ variation $\delta T$ is determined by the direction of the electric field $E$. 

More precisely, $E/T$ ratio is a constant in the conditions of electrocaloric effect. Thus, $\delta (E/T)$ is zero, which gives $\delta T = (T/E) \delta E$. In our experiment,  we measure only  $T\delta E$ component which is proportional to $(I_{ac})^2$ due to the $2\omega$ lock-in detection technique, while the in-plane electric field  $E\sim I_{dc} \sim V_{dc}$. It gives exactly $\nabla T \sim (I_{ac})^2/I_{dc}$ dependence, which we observe in Figs.~\ref{Vxy} and ~\ref{Vxy_II sample}. 

From the $V_{xy}^{2\omega}$ values within $\pm$0.3~mV one can estimate~\cite{TE} maximum temperature variation as $\Delta T \approx \pm0.5$~K. This value well corresponds for the known one (mostly 2-5~K) in insulating ferroelectric crystals~\cite{BTSn} or thin films~\cite{ECE,ferrECE,PZT}. 
Thus, we not only demonstrate possibility to create electrocaloric effect by current-induced electric field in conducting ferroelectric crystals, but also obtain competitive values of the effect.

\section{Conclusion}
As a conclusion, we experimentally investigate  thermoelectric response of a 100~nm thick SnSe single crystal flake under the current-induced dc electric field. Thermoelectric response appears as a second-harmonic transverse voltage $V_{xy}^{2\omega}$, which reflects temperature gradient across the sample due to the Joule heating by harmonic ac excitation current $I_{ac}$.  In addition to   strongly non-monotonous dependence  $V_{xy}^{2\omega}$, we observe that  dc field direction controls the sign of the temperature gradient in the SnSe flake.  We provide arguments, that electrocaloric effect is the mostly probable reason for  the results obtained. Thus, our experiment can be understood as  demonstration  of the possibility to induce electrocaloric effect by in-plane electric field in conducting ferroelectric crystals and to detect it by thermoelectric response.

\section{Acknowledgement}

We wish to thank V.T. Dolgopolov for fruitful discussions, and S.S~Khasanov for X-ray sample characterization.
We gratefully acknowledge financial support partially by the RFBR  (project No.~19-02-00203), and RF
State task.

\end{document}